\documentclass[sigconf]{acmart}

\usepackage[font=small]{caption}
\usepackage[font=small]{subcaption}
\usepackage{amsmath}
\usepackage{gensymb}

\usepackage{wrapfig}
\usepackage[keeplastbox]{flushend}

\usepackage{booktabs}
\usepackage{tabularx}
\newcolumntype{b}{X}
\newcolumntype{s}{>{\hsize=.4\hsize}X}

\usepackage[inline]{enumitem}
\usepackage{xspace}

\usepackage{dblfloatfix}    

\usepackage{soul}

\newlist{romaninline}{enumerate*}{1}
\setlist[romaninline]{label=(\roman*)}

\usepackage[acronyms,nonumberlist,nopostdot,nomain,nogroupskip]{glossaries}
\newacronym{quic}{QUIC}{Quick UDP Internet Connections}
\newacronym{3gpp}{3GPP}{3rd Generation Partnership Project}
\newacronym{adc}{ADC}{Analog to Digital Converter}
\newacronym{5g}{5G}{5th Generation}
\newacronym{aimd}{AIMD}{Additive Increase Multiplicative Decrease}
\newacronym{am}{AM}{Acknowledged Mode}
\newacronym{amc}{AMC}{Adaptive Modulation and Coding}
\newacronym{aqm}{AQM}{Active Queue Management}
\newacronym{awgn}{AGWN}{Additive White Gaussian Noise}
\newacronym{balia}{BALIA}{Balanced Link Adaptation}
\newacronym{bdp}{BDP}{Bandwidth-Delay Product}
\newacronym{bf}{BF}{Beamforming}
\newacronym{cc}{CC}{Congestion Control}
\newacronym{cdf}{CDF}{Cumulative Distribution Function}
\newacronym{ci}{CI}{Close-in free space reference}
\newacronym{cn}{CN}{Core Network}
\newacronym{cqi}{CQI}{Channel Quality Information}
\newacronym{cp}{CP}{Control Plane}
\newacronym{csirs}{CSI-RS}{Channel State Information - Reference Signal}
\newacronym{dc}{DC}{Dual Connectivity}
\newacronym{dce}{DCE}{Direct Code Execution}
\newacronym{dci}{DCI}{Downlink Control Information}
\newacronym{dl}{DL}{Downlink}
\newacronym{dmr}{DMR}{Deadline Miss Ratio}
\newacronym{dmrs}{DMRS}{DeModulation Reference Signal}
\newacronym{e2e}{E2E}{End-to-End}
\newacronym{ecn}{ECN}{Explicit Congestion Notification}
\newacronym{edf}{EDF}{Earliest Deadline First}
\newacronym{enb}{eNB}{evolved Node Base}
\newacronym{epc}{EPC}{Evolved Packet Core}
\newacronym{es}{ES}{Edge Server}
\newacronym{fdma}{FDMA}{Frequency Division Multiple Access}
\newacronym{fdd}{FDD}{Frequency Division Duplexing}
\newacronym[firstplural=Radio Access Technologies (RATs)]{rat}{RAT}{Radio Access Technology}
\newacronym{fs}{FS}{Fast Switching}
\newacronym{ftp}{FTP}{File Transfer Protocol}
\newacronym{gnb}{gNB}{Next Generation Node Base}
\newacronym{harq}{HARQ}{Hybrid Automatic Repeat reQuest}
\newacronym{hetnet}{HetNet}{Heterogeneous Network}
\newacronym{hh}{HH}{Hard Handover}
\newacronym{hol}{HOL}{Head-of-Line}
\newacronym{ia}{IA}{Initial Access}
\newacronym{imt}{IMT}{International Mobile Telecommunication}
\newacronym{iot}{IoT}{Internet of Things}
\newacronym{los}{LOS}{Line of Sight}
\newacronym{lte}{LTE}{Long Term Evolution}
\newacronym{m2m}{M2M}{Machine to Machine}
\newacronym{mac}{MAC}{Medium Access Control}
\newacronym{mc}{MC}{Multi-Connectivity}
\newacronym{mcs}{MCS}{Modulation and Coding Scheme}
\newacronym{mec}{MEC}{Mobile Edge Cloud}
\newacronym{mi}{MI}{Mutual Information}
\newacronym{mimo}{MIMO}{Multiple Input, Multiple Output}
\newacronym{mmwave}{mmWave}{millimeter wave}
\newacronym{mr}{MR}{Maximum Rate}
\newacronym{mss}{MSS}{Maximum Segment Size}
\newacronym{mtd}{MTD}{Machine-Type Device}
\newacronym{mtu}{MTU}{Maximum Transmission Unit}
\newacronym{nsf}{NSF}{National Science Foundation}
\newacronym{nfv}{NFV}{Network Function Virtualization}
\newacronym{nlos}{NLOS}{Non Line of Sight}
\newacronym{nr}{NR}{New Radio}
\newacronym{ofdm}{OFDM}{Orthogonal Frequency Division Multiplexing}
\newacronym{pdcch}{PDCCH}{Physical Downlonk Control Channel}
\newacronym{pdcp}{PDCP}{Packet Data Convergence Protocol}
\newacronym{pdsch}{PDSCH}{Physical Downlink Shared Channel}
\newacronym{pdu}{PDU}{Packet Data Unit}
\newacronym{pf}{PF}{Proportional Fair}
\newacronym{pgw}{PGW}{Packet Gateway}
\newacronym{phy}{PHY}{Physical}
\newacronym{pbch}{PBCH}{Physical Broadcast Channel}
\newacronym[plural=\gls{mme}s,firstplural=Mobility Management Entities (MMEs)]{mme}{MME}{Mobility Management Entity}
\newacronym{prb}{PRB}{Physical Resource Block}
\newacronym{pss}{PSS}{Primary Synchronization Signal}
\newacronym{pucch}{PUCCH}{Physical Uplink Control Channel}
\newacronym{pusch}{PUSCH}{Physical Uplink Shared Channel}
\newacronym{rach}{RACH}{Random Access Channel}
\newacronym{ran}{RAN}{Radio Access Network}
\newacronym{red}{RED}{Random Early Detection}
\newacronym{rf}{RF}{Radio Frequency}
\newacronym{rlc}{RLC}{Radio Link Control}
\newacronym{rlf}{RLF}{Radio Link Failure}
\newacronym{rrc}{RRC}{Radio Resource Control}
\newacronym{rrm}{RRM}{Radio Resource Management}
\newacronym{rr}{RR}{Round Robin}
\newacronym{rs}{RS}{Remote Server}
\newacronym{rsrp}{RSRP}{Reference Signal Received Power}
\newacronym{rss}{RSS}{Received Signal Strength}
\newacronym{rtt}{RTT}{Round Trip Time}
\newacronym{rw}{RW}{Receive Window}
\newacronym{rx}{RX}{Receiver}
\newacronym{sa}{SA}{standalone}
\newacronym{sack}{SACK}{Selective Acknowledgment}
\newacronym{sap}{SAP}{Service Access Point}
\newacronym{sch}{SCH}{Secondary Cell Handover}
\newacronym{scoot}{SCOOT}{Split Cycle Offset Optimization Technique}
\newacronym{sdma}{SDMA}{Spatial Division Multiple Access}
\newacronym{sinr}{SINR}{Signal to Interference plus Noise Ratio}
\newacronym{sm}{SM}{Saturation Mode}
\newacronym{snr}{SNR}{Signal to Noise Ratio}
\newacronym{son}{SON}{Self-Organizing Network}
\newacronym{ss}{SS}{Synchronization Signal}
\newacronym{srs}{SRS}{Sounding Reference Signal}
\newacronym{sss}{SSS}{Secondary Synchronization Signal}
\newacronym{tb}{TB}{Transport Block}
\newacronym{tcp}{TCP}{Transmission Control Protocol}
\newacronym{tdd}{TDD}{Time Division Duplexing}
\newacronym{tdma}{TDMA}{Time Division Multiple Access}
\newacronym{tfl}{TfL}{Transport for London}
\newacronym{thz}{THz}{Terahertz}
\newacronym{tm}{TM}{Transparent Mode}
\newacronym{trp}{TRP}{Transmitter Receiver Pair}
\newacronym{tti}{TTI}{Transmission Time Interval}
\newacronym{ttt}{TTT}{Time-to-Trigger}
\newacronym{tx}{TX}{Transmitter}
\newacronym{ue}{UE}{User Equipment}
\newacronym{ul}{UL}{Uplink}
\newacronym{uml}{UML}{Unified Modeling Language}
\newacronym{um}{UM}{Unacknowledged Mode}
\newacronym{utc}{UTC}{Urban Traffic Control}
\newacronym{vm}{VM}{Virtual Machine}
\newacronym{rsrq}{RSRQ}{Reference Signal Received Quality}
\newacronym{rssi}{RSSI}{Received Signal Strength Indicator}
\newacronym{crs}{CRS}{Cell Reference Signal}
\newacronym{comp}{CoMP}{Coordinated Multi-Point}
\newacronym{cran}{C-RAN}{Cloud \acrlong{ran}}
\newacronym{ca}{CA}{Carrier Aggregation}
\newacronym{cco}{CC}{Carrier Component}
\newacronym{nsa}{NSA}{Non Stand Alone}
\newacronym{embb}{eMBB}{Enhanced Mobility Broadband}
\newacronym{bsr}{BSR}{Buffer Status Report}
\newacronym{srb}{SRB}{Service Radio Bearer}
\newacronym{scm}{SCM}{Spatial Channel Model}
\newacronym{sctp}{SCTP}{Stream Control Transmission Protocol}
\newacronym{mptcp}{MPTCP}{Multi-path TCP}
\newacronym{ietf}{IETF}{Internet Engineering Task Force}
\newacronym{os}{OS}{Operating System}
\newacronym{tls}{TLS}{Transport Layer Security}
\newacronym{rfc}{RFC}{Request for Comments}
\newacronym{http}{HTTP}{HyperText Transfer Protocol}
\newacronym{nat}{NAT}{Network Address Translation}
\newacronym{api}{API}{Application Programming Interface}
\newacronym{rto}{RTO}{Retransmission Timeout}
\newacronym{psc}{PSC}{Public Safety Communication}
\newacronym{rpgm}{RPGM}{Reference Point Group Mobility}
\newacronym{ic}{IC}{Incident Command}
\newacronym{rsu}{RSU}{Road Side Unit}
\newacronym{uav}{UAV}{Unmanned Aerial Vehicle}
\newacronym{usa}{U.S.}{United States}
\newacronym{vr}{VR}{Virtual Reality}
\newacronym{iab}{IAB}{Integrated Access and Backhaul}
\newacronym{wlan}{WLAN}{Wireless Local Area Network}
\newacronym{cots}{COTS}{Commercial Off-the-Shelf}
\newacronym{fpga}{FPGA}{Field Programmable Gate Array}
\newacronym{rcn}{RCN}{Research Coordination Network}
\newacronym{abg}{ABG}{Alpha-Beta-Gamma}
\newacronym{fi}{FI}{Floating Intercept}
\newacronym{uas}{UAS}{Unmanned Aerial System}
\newacronym{gps}{GPS}{Global Positioning System}
\newacronym{a2g}{A2G}{air-to-ground}
\newacronym{a2a}{A2A}{air-to-air}
\newacronym{uma}{UMa}{Urban Macro}
\newacronym{umi}{UMi}{Urban Micro}
\newacronym{rma}{RMa}{Rural Macro}
\newacronym{inoo}{InOo}{Indoor Open Office}
\newacronym{ple}{PLE}{path loss exponent}
\newacronym{aoa}{AoA}{Angle of Arrival}
\newacronym{aod}{AoD}{Angle of Departure}
\newacronym{toa}{ToA}{Time of Arrival}
\newacronym{mpc}{MPC}{Multi-path Component}
\newacronym{cir}{CIR}{Channel Impulse Response}
\newacronym{rt}{RT}{Ray-tracing}
\newacronym{tc}{TC}{Time Cluster}
\newacronym{sl}{SL}{Spatial Lobe}
\newacronym{6g}{6G}{Sixth Generation}
\newacronym{ns3}{ns-3}{Network Simulator 3}
\newacronym{fsc}{FS}{Fully Stochastic}
\newacronym{hbc}{HB}{Hybrid}
\newacronym{hpbw}{HPBW}{Half Power Beamwidth}

\usepackage{tikz}
\usepackage{pgfplots}
\pgfplotsset{compat=newest} 
\pgfplotsset{plot coordinates/math parser=false} 
\newlength\fheight
\newlength\fwidth
\usetikzlibrary{plotmarks,patterns,decorations.pathreplacing,backgrounds,calc,arrows,arrows.meta,spy,matrix,decorations.markings}
\usepgfplotslibrary{patchplots,groupplots}
\usepackage{tikzscale}

\tikzstyle{startstop} = [rectangle, rounded corners, minimum width=2cm, minimum height=0.5cm,text centered, draw=black]
\tikzstyle{io} = [trapezium, trapezium left angle=70, trapezium right angle=110, minimum width=3cm, minimum height=1cm, text centered, draw=black]
\tikzstyle{process} = [rectangle, minimum width=2cm, minimum height=0.5cm, text centered, draw=black, align=center]
\tikzstyle{decision} = [ellipse, minimum width=2cm, minimum height=1cm, text centered, draw=black]
\tikzstyle{arrow} = [thick,<->,>=stealth]
\tikzstyle{line} = [thick,>=stealth]
\tikzstyle{lineDashed} = [thick,>=stealth,dashed]
\tikzstyle{darrow} = [thick,<->,>=stealth]
\tikzstyle{sarrow} = [thick,->,>=stealth]
\tikzstyle{larrow} = [line width=0.1mm,dashdotted,<->,>=stealth]
\tikzstyle{vecArrow} = [thick, decoration={markings,mark=at position
   1 with {\arrow[semithick, fill=white]{open triangle 60}}},
   double distance=1.4pt, shorten >= 5.5pt,
   preaction = {decorate},
   postaction = {draw,line width=1.4pt, white,shorten >= 4.5pt}]
\tikzstyle{innerWhite} = [semithick, white,line width=1.4pt, shorten >= 4.5pt]

\definecolor{SchoolColor}{RGB}{0.71, 0, 0.106}
\definecolor{chaptergrey}{rgb}{0.61, 0, 0.09} 
\definecolor{midgrey}{rgb}{0.4, 0.4, 0.4}
\definecolor{chaptergreen}{rgb}{0.09, 0.612, 0}
\definecolor{chapterpurple}{rgb}{0.522, 0, 0.612}
\definecolor{chapterlightgreen}{rgb}{0, 0.612, 0.522}

\makeatletter
\def\grd@save@target#1{%
  \def\grd@target{#1}}
\def\grd@save@start#1{%
  \def\grd@start{#1}}
\tikzset{
  grid with coordinates/.style={
    to path={%
      \pgfextra{%
        \edef\grd@@target{(\tikztotarget)}%
        \tikz@scan@one@point\grd@save@target\grd@@target\relax
        \edef\grd@@start{(\tikztostart)}%
        \tikz@scan@one@point\grd@save@start\grd@@start\relax
        \draw[minor help lines] (\tikztostart) grid (\tikztotarget);
        \draw[major help lines] (\tikztostart) grid (\tikztotarget);
        \grd@start
        \pgfmathsetmacro{\grd@xa}{\the\pgf@x/1cm}
        \pgfmathsetmacro{\grd@ya}{\the\pgf@y/1cm}
        \grd@target
        \pgfmathsetmacro{\grd@xb}{\the\pgf@x/1cm}
        \pgfmathsetmacro{\grd@yb}{\the\pgf@y/1cm}
        \pgfmathsetmacro{\grd@xc}{\grd@xa + \pgfkeysvalueof{/tikz/grid with coordinates/major step x}}
        \pgfmathsetmacro{\grd@yc}{\grd@ya + \pgfkeysvalueof{/tikz/grid with coordinates/major step y}}
        \foreach \x in {\grd@xa,\grd@xc,...,\grd@xb}
        \node[anchor=north] at (\x,\grd@ya) {\pgfmathprintnumber{\x}};
        \foreach \y in {\grd@ya,\grd@yc,...,\grd@yb}
        \node[anchor=east] at (\grd@xa,\y) {\pgfmathprintnumber{\y}};
      }
    }
  },
  minor help lines/.style={
    help lines,
    gray,
    line cap =round,
    xstep=\pgfkeysvalueof{/tikz/grid with coordinates/minor step x},
    ystep=\pgfkeysvalueof{/tikz/grid with coordinates/minor step y}
  },
  major help lines/.style={
    help lines,
    line cap =round,
    line width=\pgfkeysvalueof{/tikz/grid with coordinates/major line width},
    xstep=\pgfkeysvalueof{/tikz/grid with coordinates/major step x},
    ystep=\pgfkeysvalueof{/tikz/grid with coordinates/major step y}
  },
  grid with coordinates/.cd,
  minor step x/.initial=.5,
  minor step y/.initial=.2,
  major step x/.initial=1,
  major step y/.initial=1,
  major line width/.initial=1pt,
}
\makeatother

\title{Full-stack Comparison of Channel Models for Networks Above 100 GHz in an Indoor Scenario}

\author{Amir Ashtari Gargari}
\affiliation{
\institution{Department of\\Information Engineering\\University of Padova\\Padova, Italy}}
\email{amirashtari@dei.unipd.it}

\author{Michele Polese}
\affiliation{
\institution{Institute for the Wireless\\Internet of Things\\Northeastern University\\Boston, MA, USA}}
\email{m.polese@northeastern.edu}

\author{Michele Zorzi}
\affiliation{
\institution{Department of\\Information Engineering\\University of Padova\\Padova, Italy}}
\email{zorzi@dei.unipd.it}

\begin{abstract}
    The \gls{6g} of mobile networks is expected to use carrier frequencies in the spectrum above 100 GHz, to satisfy the demands for higher data rates and bandwidth of future digital applications. The development of networking solutions at such high frequencies is challenged by the harsh propagation environment, and by the need for directional communications and signal processing at high data rates. A fundamental step in defining and developing wireless networks above 100 GHz is given by an accurate performance evaluation. For simulations, this strongly depends on the accuracy of the modeling of the channel and of the interaction with the higher layers of the stack. This paper introduces the implementation of two recently proposed channel models (based on ray tracing and on a fully stochastic model) for the 140 GHz band for the ns-3 TeraSim module, which enables simulation of macro wireless networks in the sub-terahertz and terahertz spectrum. We also compare the two channel models with full-stack simulations in an indoor scenario, highlighting differences and similarities in how they interact with the protocol stack and antenna model of TeraSim.
\end{abstract}

\keywords{Terahertz, Cellular Networks, mmWave, channel modeling.}

\ccsdesc[500]{Networks~Network performance evaluation}
\ccsdesc[500]{Networks~Mobile networks}

\acmYear{2022}\copyrightyear{2022}
\setcopyright{acmlicensed}
\acmConference[mmNets '21]{5th ACM Workshop on Millimeter-Wave and Terahertz Networks and Sensing Systems}{January 31--February 4, 2022}{New Orleans, LA, USA}
\acmBooktitle{5th ACM Workshop on Millimeter-Wave and Terahertz Networks and Sensing Systems (mmNets '21), January 31--February 4, 2022, New Orleans, LA, USA}
\acmPrice{15.00}
\acmDOI{10.1145/3477081.3481677}
\acmISBN{978-1-4503-8699-9/22/01}

\begin{document}

\fancyhead{}

\maketitle

\glsresetall

\section{Introduction}
\label{sec:intro}

The society and the economy are becoming more and more reliant on wireless connectivity as the backbone for digital services. Mobile connected devices enable human-to-human communications, remote machine monitoring, control, actuation, as part of the Industry 4.0 paradigm, improve autonomous driving through high-speed data exchange, and empower remote health solutions~\cite{giordani20196g}. As the number of services and users of the wireless spectrum increases, so does the need for increased bandwidth and higher data rates. The \gls{5g} of mobile networks features the first \gls{3gpp} mobile \gls{ran} (i.e., \gls{3gpp} NR) to support frequencies in the lower \gls{mmwave} range, i.e., below 52.6 GHz, with extensions to 71 GHz being considered for future \gls{3gpp} releases~\cite{3gpp.38.300,38808}. 

Nowadays, \gls{5g} is being commercially deployed, and the research and industrial community have shifted their focus toward the development of \gls{6g} solutions~\cite{giordani20196g}. \gls{6g} networks will support carrier frequencies in the spectrum above 100 GHz, in the lower part of the terahertz bands, to fulfill the quest for additional bandwidth~\cite{akyildiz2014terahertz}. For instance, the IEEE 802.15.3d standard has been developed to operate in sub-terahertz frequency bands with up to 69 GHz of bandwidth, to support new applications such as \gls{vr} and telepresence through 3D holograms~\cite{ieee.subterahertz,giordani20196g}.  Notably, the spectrum available above 100 GHz is expected to enable terabit-per-second links, which could be also used for high-capacity backhaul, and to facilitate the integration of new paradigms in communications and sensing~\cite{sariedden2020next}.

\begin{table*}[t]
    \centering
    \small
    \begin{tabularx}{\linewidth}{>{\hsize=.4\hsize}X>{\hsize=.4\hsize}X>{\hsize=.8\hsize}X>{\hsize=.4\hsize}XX}
        \toprule
         Channel Model & Frequency & Modeling approach & LOS/NLOS & Characteristics \\\midrule
         TeraSim~\cite{hossain2018terasim} & 0.1-10 THz & Physics-based & LOS propagation & Frequency-selective model for propagation at terahertz \\\rule{0pt}{2.5ex}
         \gls{fsc} Channel~\cite{Statchannel} & 28 GHz, 140 GHz & \gls{scm}, measurement-based & LOS/NLOS, fading & Stochastic model, scales to randomly generated scenarios for large scale evaluations \\\rule{0pt}{2.5ex}
         \gls{hbc} Channel~\cite{chen2021channel} & 130-143 GHz & \gls{scm} based on RT and random components, measurement-based & LOS/NLOS, fading & The baseline model is given by \gls{rt}, accurate in specific, deterministic scenarios. \\

         \bottomrule
    \end{tabularx}
    \caption{Channel models for above 100 GHz wireless networks considered in this paper.}
    \label{tab:channels}
\end{table*}
\glsreset{hbc}
\glsreset{fsc}
\glsreset{scm}

The support of carriers at such high frequencies, however, comes with several challenges. For example, the propagation environment is challenging, with a high pathloss, which increases proportionally with the square of the carrier frequency, as the size of an antenna element decreases, and with additional molecular absorption in specific frequency bands. Furthermore, communications in the \gls{mmwave} and terahertz spectrum are affected by blockage, as signals do not penetrate common materials. From a networking perspective, directionality is key to improving the link budget, but requires the two endpoints of a link to coordinate to identify the best directions to communicate with each other. These issues, together with the challenging data processing at high rates, propagate throughout the whole protocol stack~\cite{polese2020toward}.

The development of solutions, algorithms, and protocols for wireless networks above 100 GHz necessitates tools for accurate and reliable performance evaluation. At the time of writing, wireless testbeds for such frequency bands are extremely expensive, due to the need for custom \gls{rf} components. They are thus limited to few nodes \cite{sen2020experimental}, and mostly focused on physical layer evaluations or channel sounding. In this context, simulations can be used to provide an end-to-end, full-stack performance evaluation platform. This has helped, for example, highlight a number of issues that arise from the interaction
between the higher layers of the protocol stack, such as TCP, and the large bandwidth, the blockage phenomena, and the directional communications typical of these frequency bands~\cite{polese2020toward,zhang2019will}. The accuracy of simulations, however, strongly depends on the accuracy of the channel model, and on how it is possible to represent its interactions with the rest of the device and protocol stack components (e.g., the directional antenna components), as also discussed in~\cite{lecci2021accuracy}. 

Recent papers have proposed measurement-based channel models for the spectrum above 100 GHz~\cite{chen2021channel,Statchannel,yuanbo2021ray}, but they have not been considered in the context of an end-to-end, full-stack deployment. Other work has focused on the development of simulation tools for terahertz frequencies, but with a \gls{los} channel model~\cite{hossain2018terasim}. This paper combines both, enabling a full-stack evaluation of wireless networks above 100 GHz with realistic channel models, which include \gls{nlos} and fading modeling. 

Notably, the contributions are twofold. First, we review channel modeling options for the spectrum above 100 GHz, and present the implementation for the ns-3 module TeraSim of two \glspl{scm}. The first is the \gls{fsc} \gls{scm} introduced in~\cite{Statchannel}, while the second is a \gls{hbc} channel model introduced in~\cite{chen2021channel}, which combines ray tracing with additional stochastic components. We selected these channel modeling strategies for implementation and comparison as they present a trade-off between generality and accuracy. Fully stochastic models can be applied in generic scenarios, but are not as precise as \gls{rt}-based channels when considering specific deployments. Conversely, \gls{rt}-based channels have a higher computational overhead, as they require the generation of \glspl{mpc} using an \gls{rt}. Additionally, \glspl{scm} can be used for performance evaluation with antenna arrays and directional communications.

The second contribution is a full-stack comparison of the two channel modeling strategies in an indoor office scenario. We investigate the interplay between the antenna and the channel modeling using UDP as transport protocol, and different source traffic configurations. Our results show that different channel models (with different spatial distribution of the \glspl{mpc}) have a different interplay with directional antenna models, which eventually result into different full-stack throughput and latency values. These results can inform the design of simulation campaigns for evaluation of future \gls{6g} networks in generic and/or specific scenarios.
Finally, we open source the channel modeling code as additional contribution to the community.\footnote{The code is available at \url{https://github.com/signetlabdei/thz-mmnets-2021}.}

The rest of the paper is organized as follows. Section~\ref{sec:soa} introduces and qualitatively compares the two channel models. Section~\ref{sec:sim} discusses the implementation for the ns-3 TeraSim module, and Section~\ref{sec:perfeval} presents the results for the numerical comparison between the channel models. Finally, Section~\ref{sec:conclusions} concludes the paper.

\section{Channel Modeling Above 100 GHz}
\label{sec:soa}

The development of \gls{5g} networks in the lower \gls{mmwave} band has been initially driven by the identification of channel models, such as those discussed in~\cite{rappaport2013millimeter,ferrand2016trends}. Similarly, the focus of channel modeling in above 100 GHz bands is shifting from the analysis of the propagation and molecular absorption~\cite{jornet2011channel} to \gls{scm} models~\cite{han2015multi,peng2020channel,moldovan2014los,han2018propagation}. These are useful for evaluations with antenna arrays, and include fading and \gls{los}/\gls{nlos} characterizations. Recent examples are the models published in~\cite{chen2021channel,Statchannel,yuanbo2021ray}. Table~\ref{tab:channels} summarizes the common characteristics and differences between the different modeling approaches considered in this paper, which we selected as representative of fully stochastic and \gls{rt}-based models, as discussed in Section~\ref{sec:intro}. In the next paragraphs, we will discuss in detail the \gls{fsc} and 
\gls{hbc} models, which will then be compared in Sec.~\ref{sec:perfeval}.

\subsection{Hybrid Channel Modeling}



The \gls{hbc} model \cite{chen2021channel} characterizes indoor sub-terahertz propagation by combining \gls{rt} techniques and stochastic methods. This channel model is built upon wideband channel measurements conducted at 130–143 GHz frequencies in a standard conference room. The \gls{hbc} model comprises \gls{mpc} clustering and matching procedures with \gls{rt} techniques to study the cluster behavior and wave propagation in the terahertz spectrum \cite{RT3D}. The \gls{rt} component makes it possible to accurately account for realistic scattering conditions in a specific scenario, while the stochastic element models random scatterers (e.g., wall texture, or small objects) that may be difficult to properly capture within the \gls{rt} scenario~\cite{lecci2020quasi}. The \gls{cir} as a function of the frequency $f$ is given by 
\begin{equation}
h_{HB}(\tau,\theta,f)= h_{RT}(\tau,\theta,f) + h_{S}(\tau,\theta,f),
\label{EQ_HB1}
\end{equation}
where $\tau$ and $\theta$ indicate the delay and azimuth \gls{aoa}, and $h_{RT}(\cdot)$ and $h_{S}(\cdot)$ are the channel components modeled through \gls{rt} and stochastic methods. $h_{RT}$ combines the Fresnel equations with the geometrical data to estimate the \gls{los} and the reflection losses from the scattering of the walls (central sub-path in \gls{rt} clusters). Conversely, $h_{S}$ models stochastically the diffraction of other sub-paths in the RT clusters, and the scattering of the reflections from additional obstacles (non-RT clusters). The \gls{cir} generated by the \gls{rt}, $h_{RT}(\tau,\theta,f)$, is represented as
\begin{equation}
\label{EQ_HB2}
\begin{split}
h_{RT}(\tau,\theta,f)& =          A_{t}(\phi_{LoS})
\alpha_{LoS}(f) \delta (\tau - \tau _{LoS})\delta (\theta - \theta _{LoS}) \\
& + \sum_{l=1}^{L_{RT}} A_t{t}(\phi_{l,0})\alpha_{l,0} (f) \delta (\tau - \tau _{l,0})\delta (\theta - \theta_{l,0}),
\end{split}
\end{equation}
where the subscript $l,0$ indicates the central path in the $l\textsuperscript{th}$ sub-path, $A_{t}(\cdot)$ represents the antenna pattern at the \gls{tx}, and $\alpha_{l,0}$, $\tau_{l,0}$, $\theta_{l,0}$ and $\phi_{l,0}$ represent amplitude gain, \gls{toa}, azimuth \gls{aoa}, and \gls{aod} vectors of the sub-path, respectively. $L_{RT}$ is the number of RT clusters. The \gls{cir} of the stochastic component, $h_S(\tau,\theta,f)$, is represented as
\begin{equation}
\label{EQ_HB3}
\begin{split}
h_{S}(\tau,\theta,f)& =       \sum_{l=1}^{L_{RT}}\sum_{p=-Q_{l}, p \neq 0}^{P_{l}}A_{t}(\phi_{l,p})
\alpha_{l,p}(f) \delta (\tau - \tau _{l,p})\delta (\theta - \theta _{l,p}) \\
& +  \sum_{q=1}^{L_{s}}\sum_{s=-T_{q}}^{S_{q}}  \alpha_{q,s} (f) \delta (\tau - \tau _{q,s})\delta (\theta - \theta _{q,s}),
\end{split}
\end{equation}
where subscripts $l,p$ and $s,q$ indicates the $p\textsuperscript{th}$ sub-path in the $l\textsuperscript{th}$ RT cluster, and the $s\textsuperscript{th}$ sub-path in the $q\textsuperscript{th}$ non-RT cluster, respectively. Notice that the antenna pattern of the \gls{tx} is not involved in the statistical CIR in \eqref{EQ_HB3}. The reason is that all statistical parameters are extracted from channel measurements  with the directional antenna \gls{tx}, so the effect of the antenna is contained in the generated amplitude.

\subsection{Fully Stochastic Modeling}


The \gls{fsc} channel model~\cite{Statchannel} is a \gls{3gpp}-like indoor spatial \gls{scm}. This model is based on an experimental channel measurement campaign at 28 GHz and 140 GHz, in an office environment. The channel model is based on the concept of \gls{tc} and \gls{sl}, which represent temporal and spatial statistics of the channel model, respectively. \glspl{tc} comprise \glspl{mpc} propagating adjacent in time, which may come from different \gls{aoa} (\gls{aod}), whereas \glspl{sl} define the main direction of arrival (departure). The directional \gls{cir} is
\begin{equation}
\label{EQ_FS1}
\begin{split}
h\textsubscript{FS}(\tau,\phi,\theta)& =       \sum_{n=1}^{N}\sum_{m=1}^{M_n} a_{n,m} e^{j\varphi_{n,m} } \delta (\tau - \tau_{n,m}) \\
& g\textsubscript{TX}(\phi -\phi_{n,m})
 g\textsubscript{RX}(\theta - \theta_{n,m}),
\end{split}
\end{equation}
where $\tau$, $\phi$, and $\theta$ are the absolute propagation delay, \gls{aod} vector, and \gls{aoa} vector, respectively. Also, $g\textsubscript{TX}$ and  $g\textsubscript{RX}$ are
the \gls{tx} and \gls{rx} complex amplitude antenna patterns, respectively.

\section{Integration in TeraSim}
\label{sec:sim}

\gls{ns3} is a discrete-event network simulation environment, which models a wide range of wireless and wired technologies, as well as the TCP/IP protocol stack and applications. TeraSim~\cite{hossain2018terasim} is an extension of \gls{ns3} for the modeling of \gls{thz} communication networks. The TeraSim module enables simulation-based testing of \gls{thz} networking protocols in the higher layers without dealing with lower layer configurations. In the following paragraphs we will briefly describe the TeraSim module, and then discuss the integration of the channel models from Sec.~\ref{sec:soa}.

\subsection{ns-3 TeraSim Module}

TeraSim is the first simulation platform for \gls{thz} communication networks that supports \gls{thz} devices' capabilities, from the protocol stack to the modeling of directional antenna patterns~\cite{xia2021link}. TeraSim considers two types of application scenarios, i.e., nano-scale scenarios, for short-range communications, and macro-scale, for traditional macro (e.g., cellular) scenarios. The module features common channel, antenna, and energy models for nano and macro applications, and separate \gls{mac} and \gls{phy} models. 

The antenna model of TeraSim is based on the directional communication scheme introduced in~\cite{xia2021link}, where alignment between mobile nodes is achieved with a rotating antenna. The model is implemented in the \textit{THzDirectionalAntenna} class, which extends the \gls{ns3} cosine antenna with the possibility to rotate. The \textit{THzDirectionalAntenna} class calculates the antenna gain based on the mobility and the RX orientation. It is possible to configure different parameters for the antenna model, including whether the antenna is static or rotates, the rotation speed and initial phase, the maximum gain and beamwidth. A beamwidth of 360 degrees models an omnidirectional antenna.

The already implemented channel model (based on~\cite{jornet2011channel}) accounts for the frequency selectivity based on different molecular absorption patterns at different frequencies of the terahertz spectrum. However, the current implementation of TeraSim only accounts for \gls{los} propagation. As the evaluation of network performance in \gls{nlos} conditions is key for a holistic evaluation of the performance of future \gls{6g} networks, we describe below the implementation of two channel models that also account for \gls{nlos} propagation.

\subsection{Integration of \gls{hbc} and \gls{fsc} Channels}



\begin{figure}[t]
    \centering
    \includegraphics[width=\linewidth]{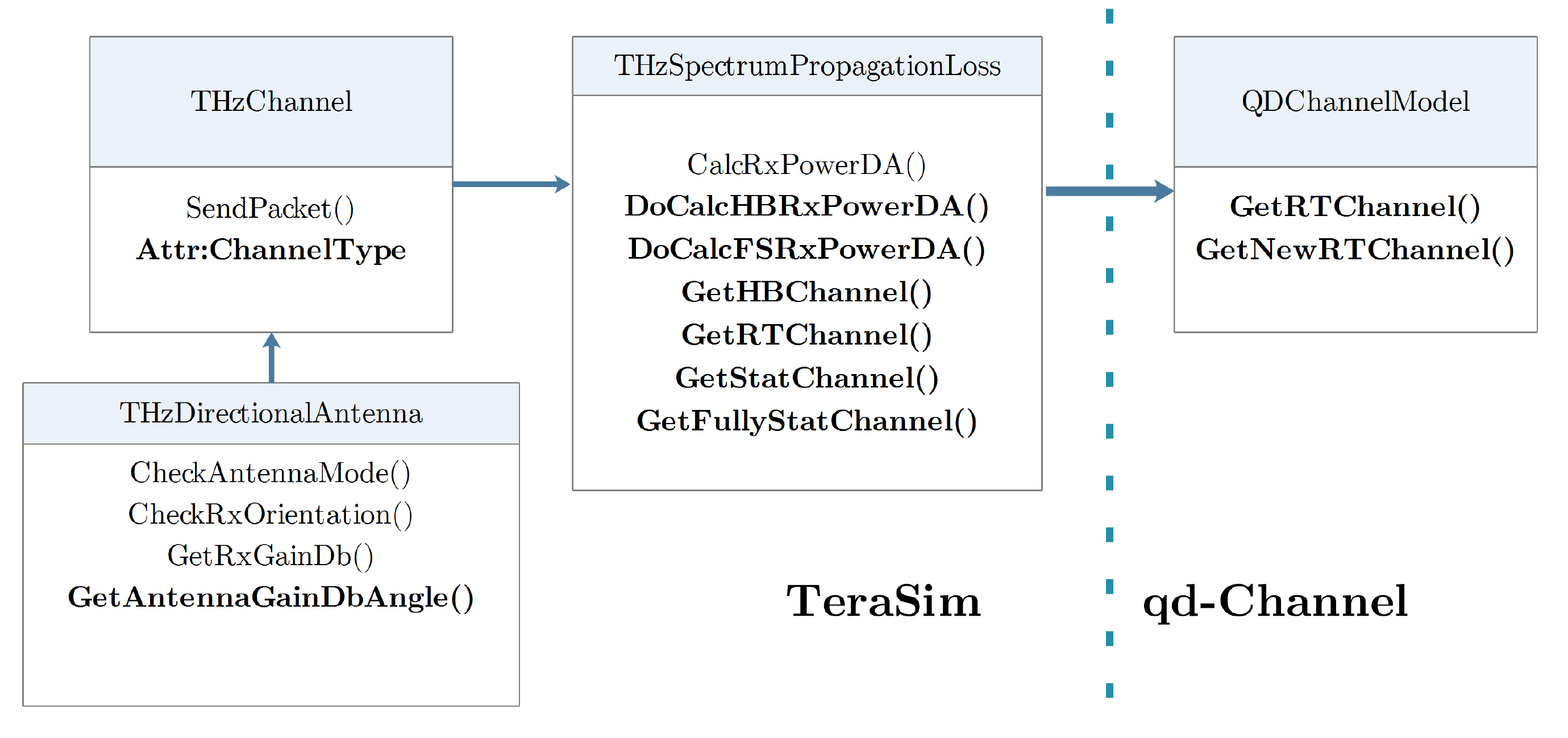}
    \caption{TeraSim classes and new channel modeling code.}
    \label{Integration}
\end{figure}

\begin{figure*}[b]
\renewcommand{\thefigure}{4}

\centering
    \begin{subfigure}[h]{0.24\textwidth}
    \centering
    \includegraphics[width=.8\textwidth]{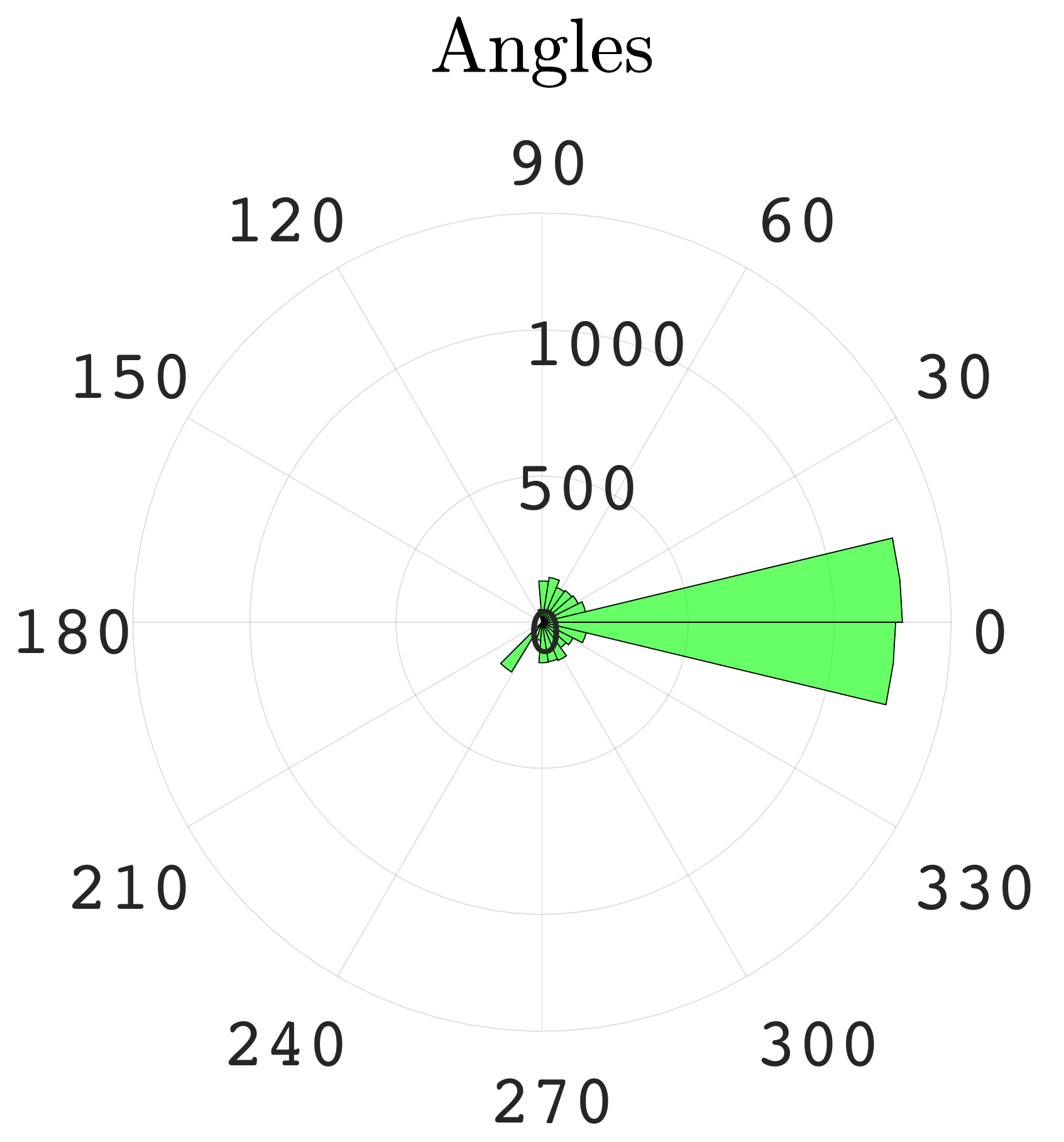}
    \caption{\gls{hbc}, \gls{los}}
    \end{subfigure}
    \begin{subfigure}[h]{0.24\textwidth}
    \centering
    \includegraphics[width=.8\textwidth]{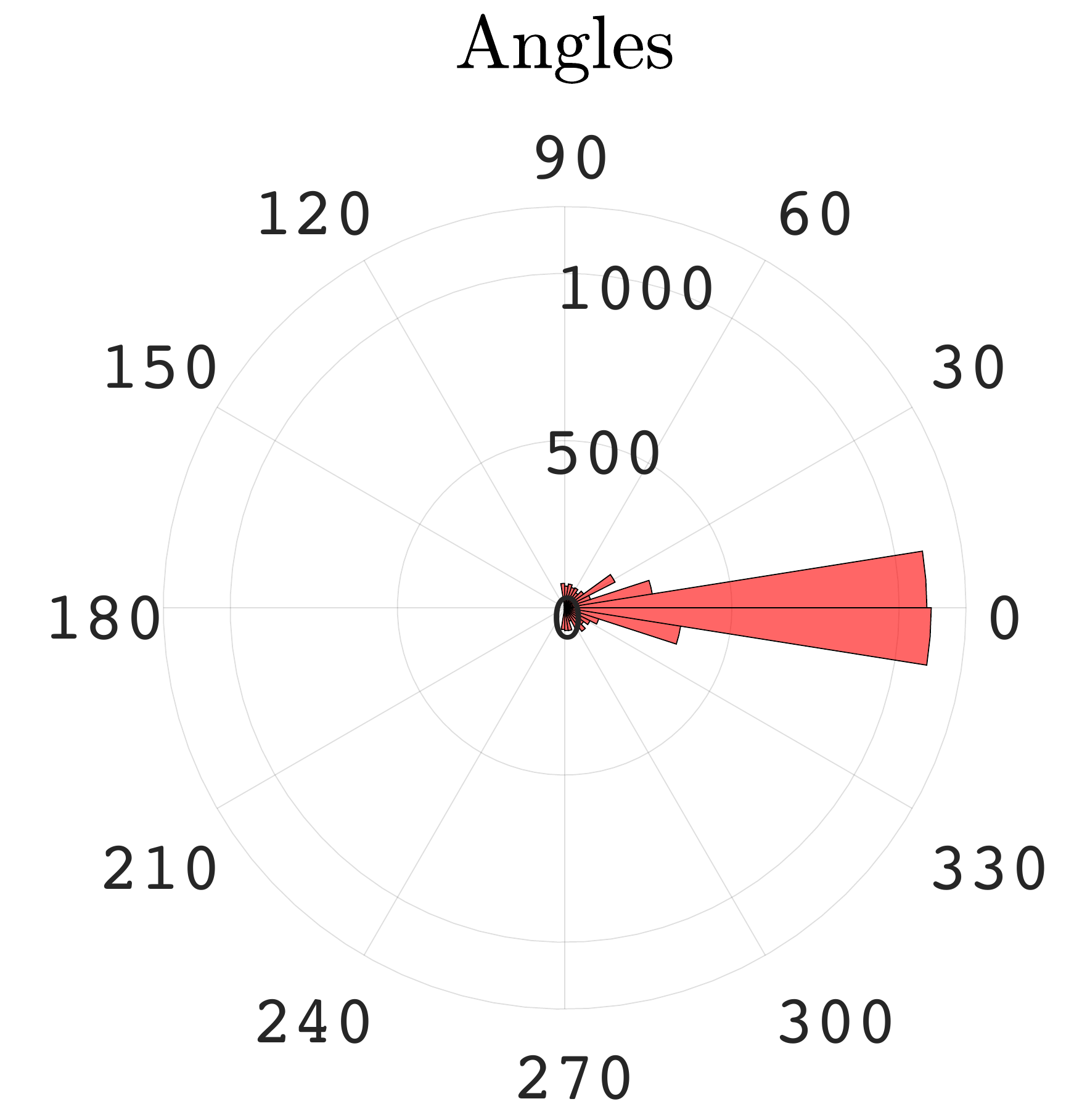}
    \caption{\gls{fsc}, \gls{los}}
    \end{subfigure}
    \begin{subfigure}[h]{0.24\textwidth}
    \centering
    \includegraphics[width=.8\textwidth]{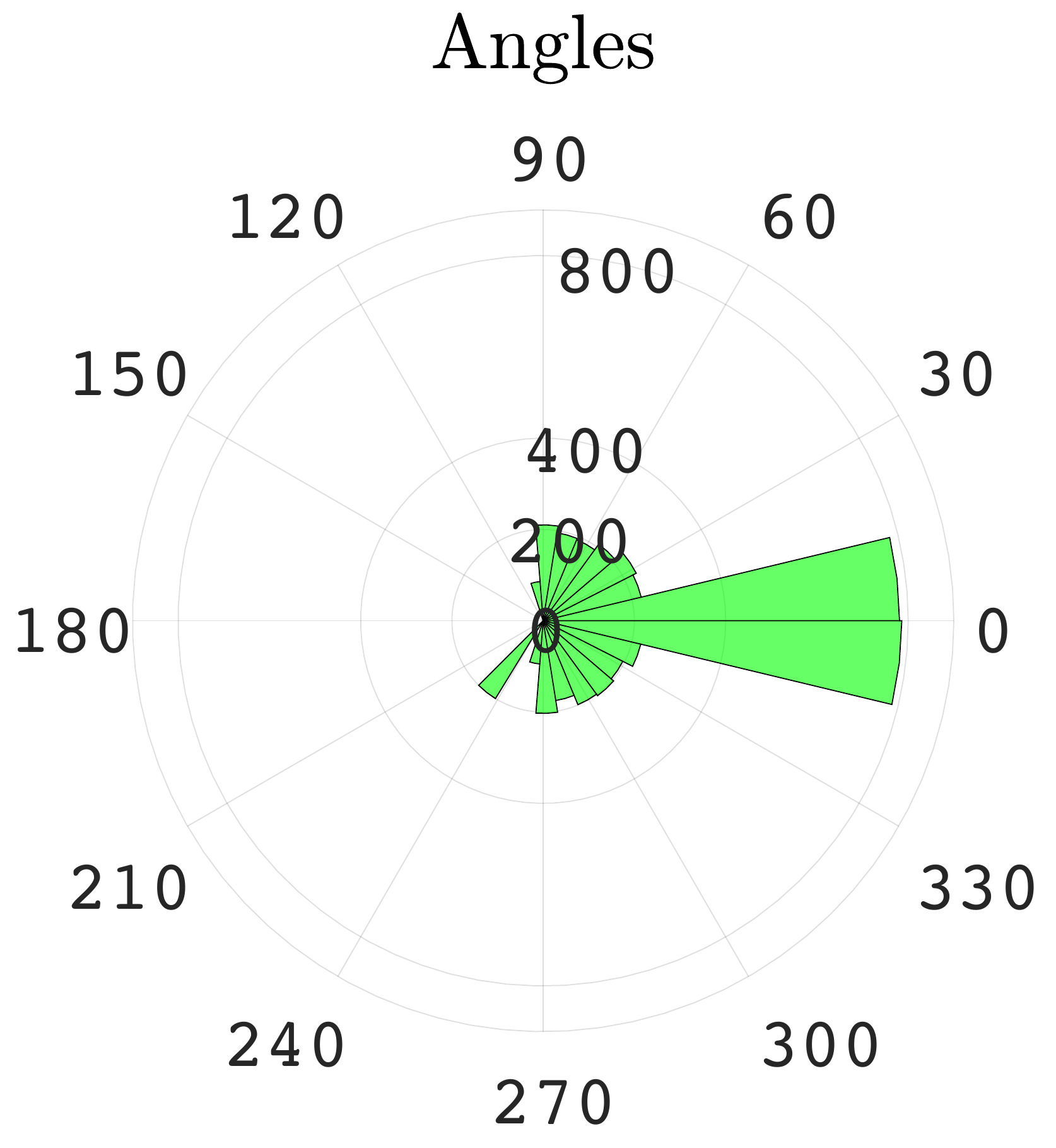}
    \caption{\gls{hbc}, \gls{nlos}}
    \end{subfigure}  
    \begin{subfigure}[h]{0.24\textwidth}
    \centering
    \includegraphics[width=.8\textwidth]{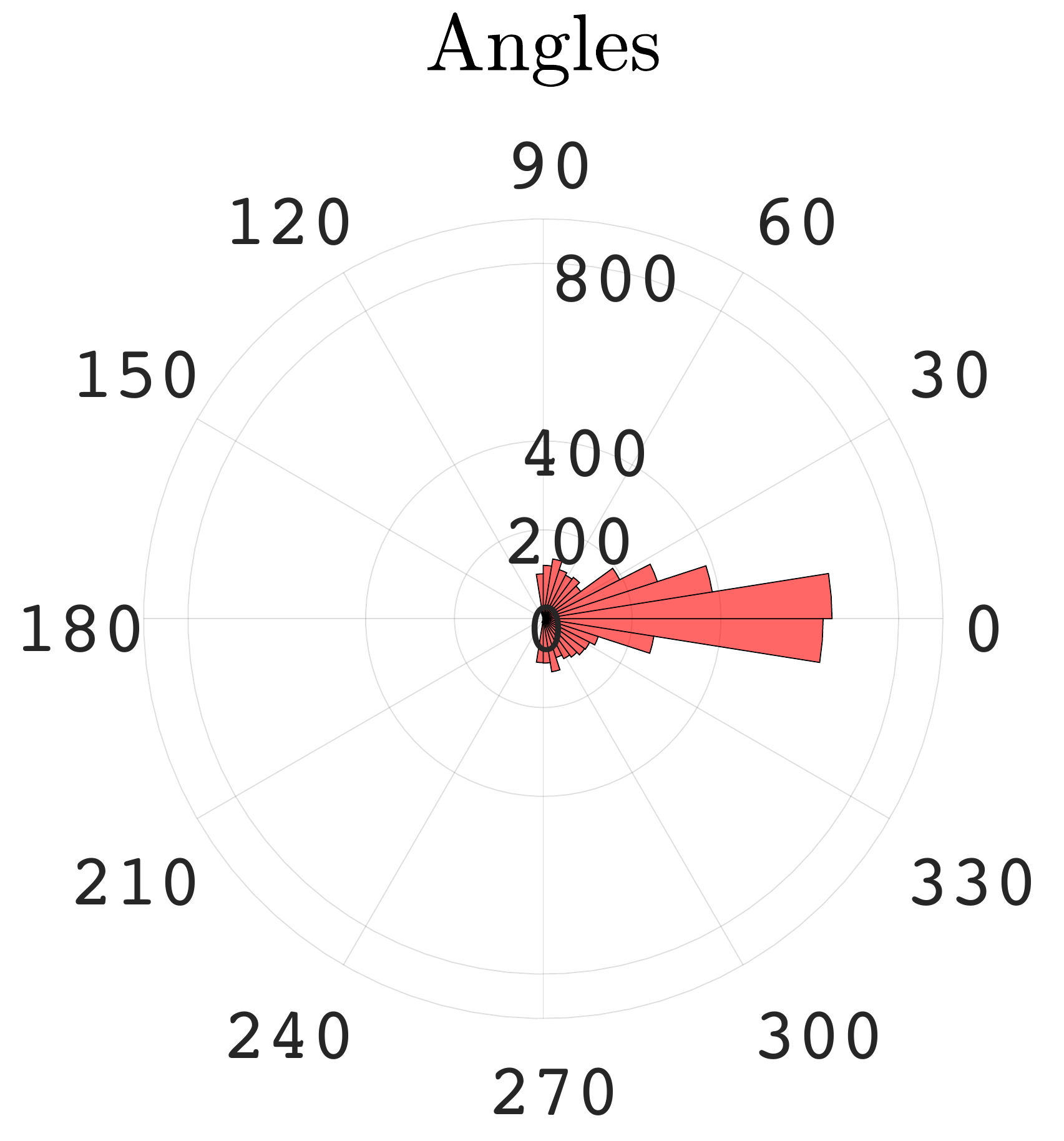}
    \caption{\gls{fsc}, \gls{nlos}}
    \end{subfigure} 
    \caption{\gls{aoa} distribution over 2000 channel realizations, for the \gls{hbc} and \gls{fsc} models, in \gls{los} and \gls{nlos}. The \gls{los} direction corresponds to 0 degrees.}
        \label{fig:aoa}
\end{figure*}

TeraSim models the waveform propagation through the wireless channel in the \textit{THzChannel} class, as shown in Figure~\ref{Integration}. The \textit{THzSpectrumValueFactory} class generates an object representing a waveform and passes it to the channel object. This first checks the orientation of the devices, and obtains the antenna gains through the \textit{THzDirectionalAntenna} module. Then, the \textit{THzSpectrum\-Propa\-gation\-Loss} class calculates the received power based on the calculated antenna gain and the \gls{cir}. The \gls{cir} is frequency selective and a function of the distance and frequency. Finally, the \textit{THzChannel} object passes the packet along with the received power to the physical layer of receivers.

Figure~\ref{Integration} summarizes the integration that we have made to support the \gls{hbc} and \gls{fsc} in TeraSim. Most of the updates have been introduced in the \textit{THzSpectrumPropagationLoss} class. Two new functions are responsible for calculating the received power for \gls{hbc} and \gls{fsc} channels, and the type of channel can be selected through the \textit{ChannelType} attribute in the \textit{THzChannel} class. Additionally, while the \gls{cir} for \gls{fsc} is generated at run-time in ns-3, the \gls{cir} for the \gls{hbc} requires the generation of the \gls{rt} \glspl{mpc} offline. For this, we use the Q-D Channel \gls{rt} tool~\cite{lecci2021accuracy,lecci2020quasi}, an open-source MATLAB-based \gls{rt} tool for the \gls{mmwave} spectrum. We updated the parameters (e.g., permittivity) in the material library and the path loss equation with those introduced in~\cite{chen2021channel,piesiewicz2007properties}. The \glspl{mpc} generated by the \gls{rt} are then loaded at run-time using the ns-3 qd-channel module.\footnote{\url{https://github.com/signetlabdei/qd-channel}}

\begin{figure}[t]
\renewcommand{\thefigure}{2}

    \centering
    \includegraphics[width=\linewidth]{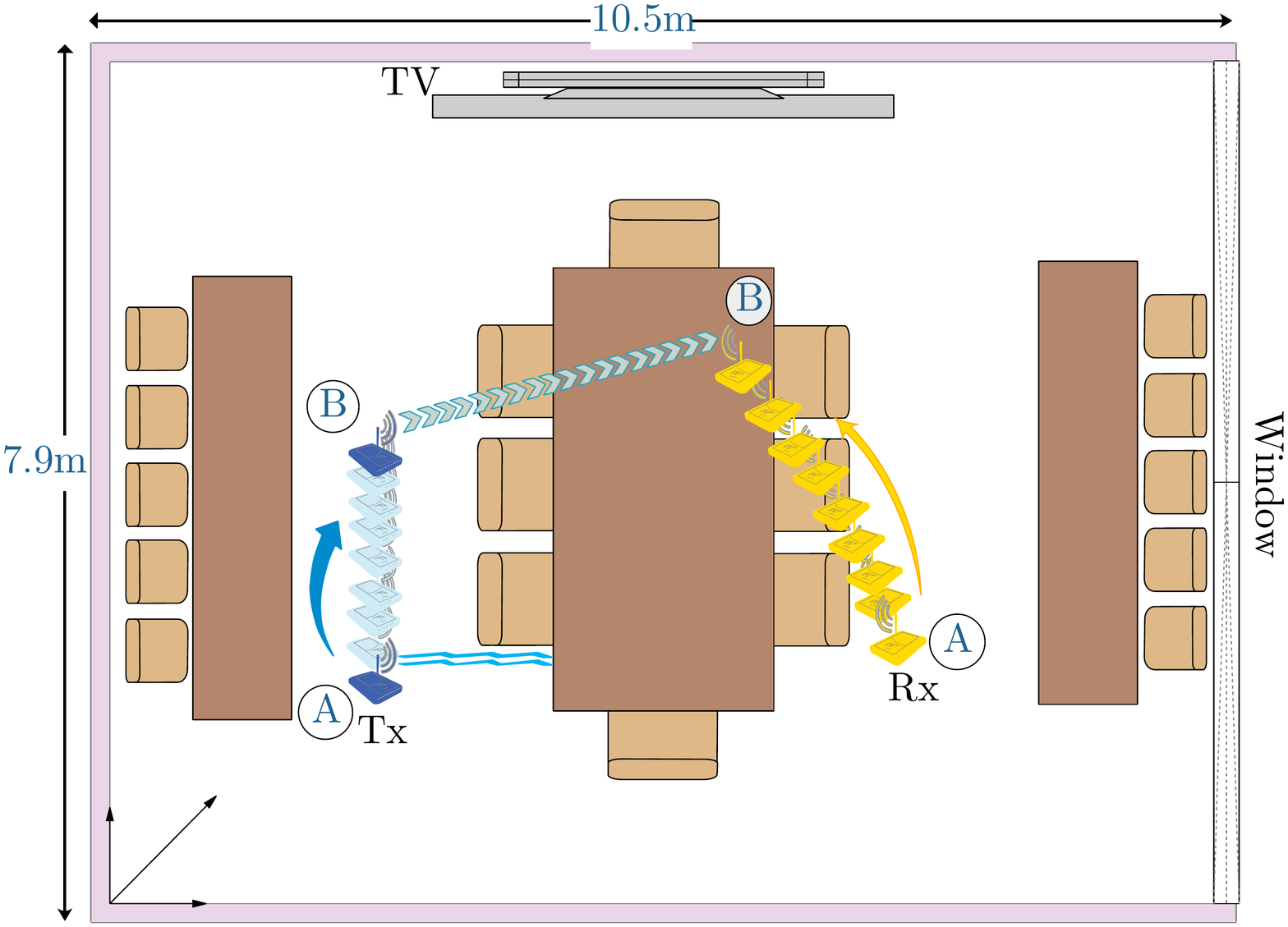}

    \caption{Simulation scenario. The \gls{tx} and \gls{rx} move from point A to point B.}
    \label{Scenario}
\end{figure}

\section{Performance Evaluation}
\label{sec:perfeval}

This section introduces a performance evaluation in an indoor scenario (Sec.~\ref{sec:scenario}), with a comparison between the results that can be obtained with the \gls{hbc} and the \gls{fsc} channel models (Sec.~\ref{sec:results}).

\subsection{Simulation Scenario}
\label{sec:scenario}

As shown in Figure~\ref{Scenario}, simulations are set in an indoor meeting room, which has the same size as that used for channel measurements in~\cite{chen2021channel}. We consider a mobility pattern for the \gls{tx} and the \gls{rx} (from point A to point B in the figure) that covers \gls{nlos} and \gls{los} conditions. The carrier frequency for the simulations is 140 GHz, with a bandwidth of 32 GHz. The \gls{tx} and \gls{rx} maximum antenna gain is 25 dBi. The \gls{hpbw} of the \gls{rx} antenna varies from 2 to 10 degrees, and the noise floor is $-160$ dBm. The application is a constant bitrate source, with source rates between 4 Gbit/s and 60 Gbit/s, and UDP as transport.

\begin{figure}[t]
\renewcommand{\thefigure}{3}

    \centering
    \setlength\fwidth{\columnwidth}
    \setlength\fheight{.5\columnwidth}
    \input{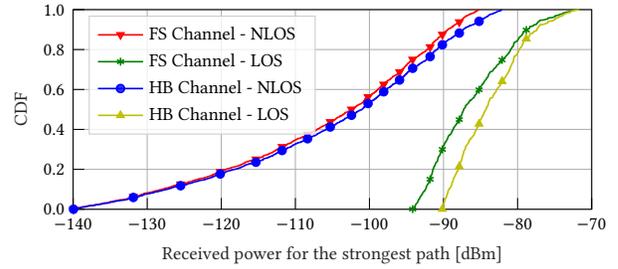}
    \setlength\abovecaptionskip{.1cm}
    \caption{\gls{cdf} of the received power (including the antenna gain) for the strongest path, beamwidth 8 degrees}
    \label{PowerCDF}
\end{figure}

\begin{figure}[t]
\centering
    \begin{subfigure}[h]{0.24\textwidth}
    \centering
    \includegraphics[width=.8\textwidth]{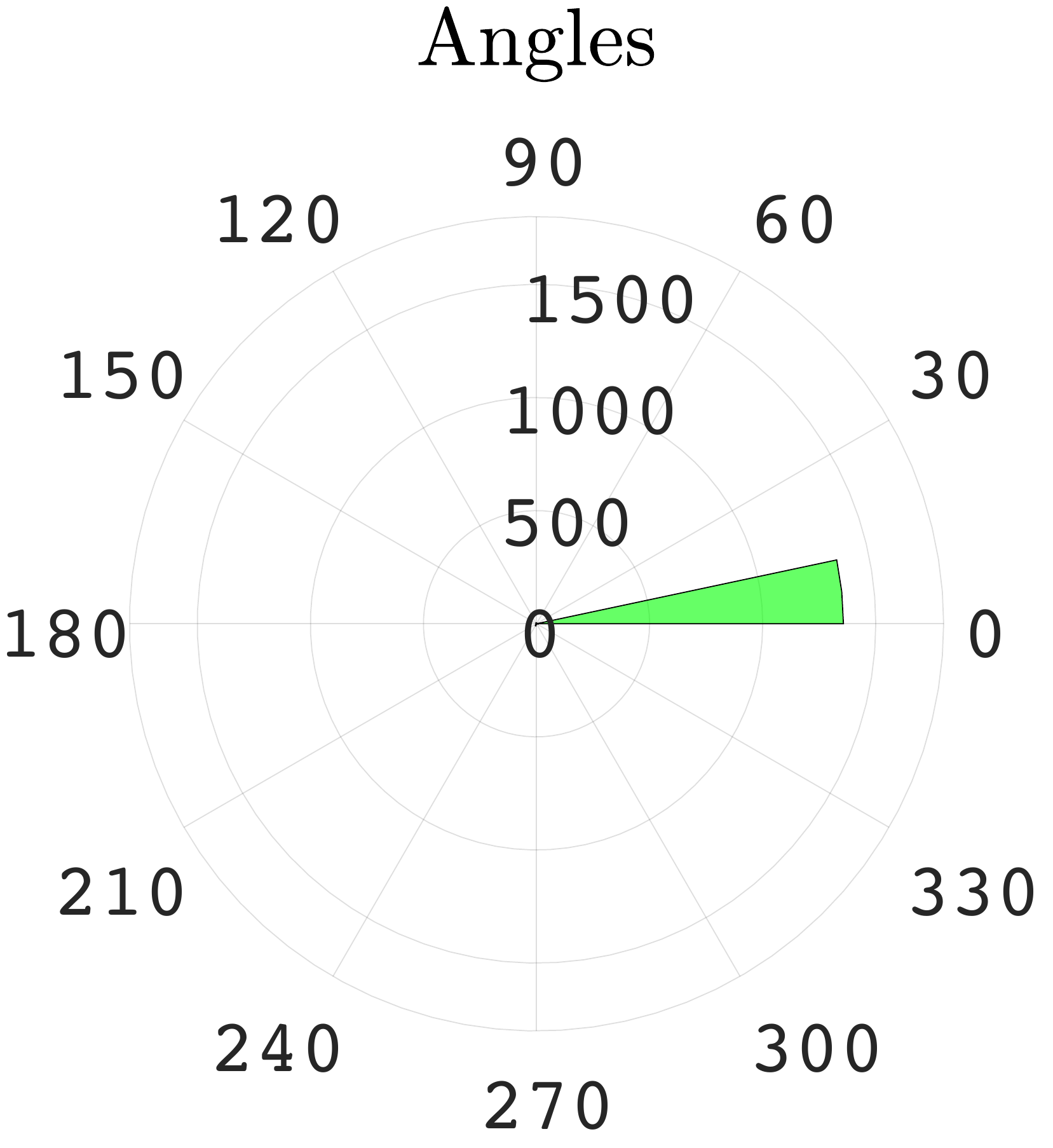}
    \caption{\gls{hbc}}
    \label{AOAMaxPowerHB}
    \end{subfigure}%
    \begin{subfigure}[h]{0.24\textwidth}
    \centering
    \includegraphics[width=.8\textwidth]{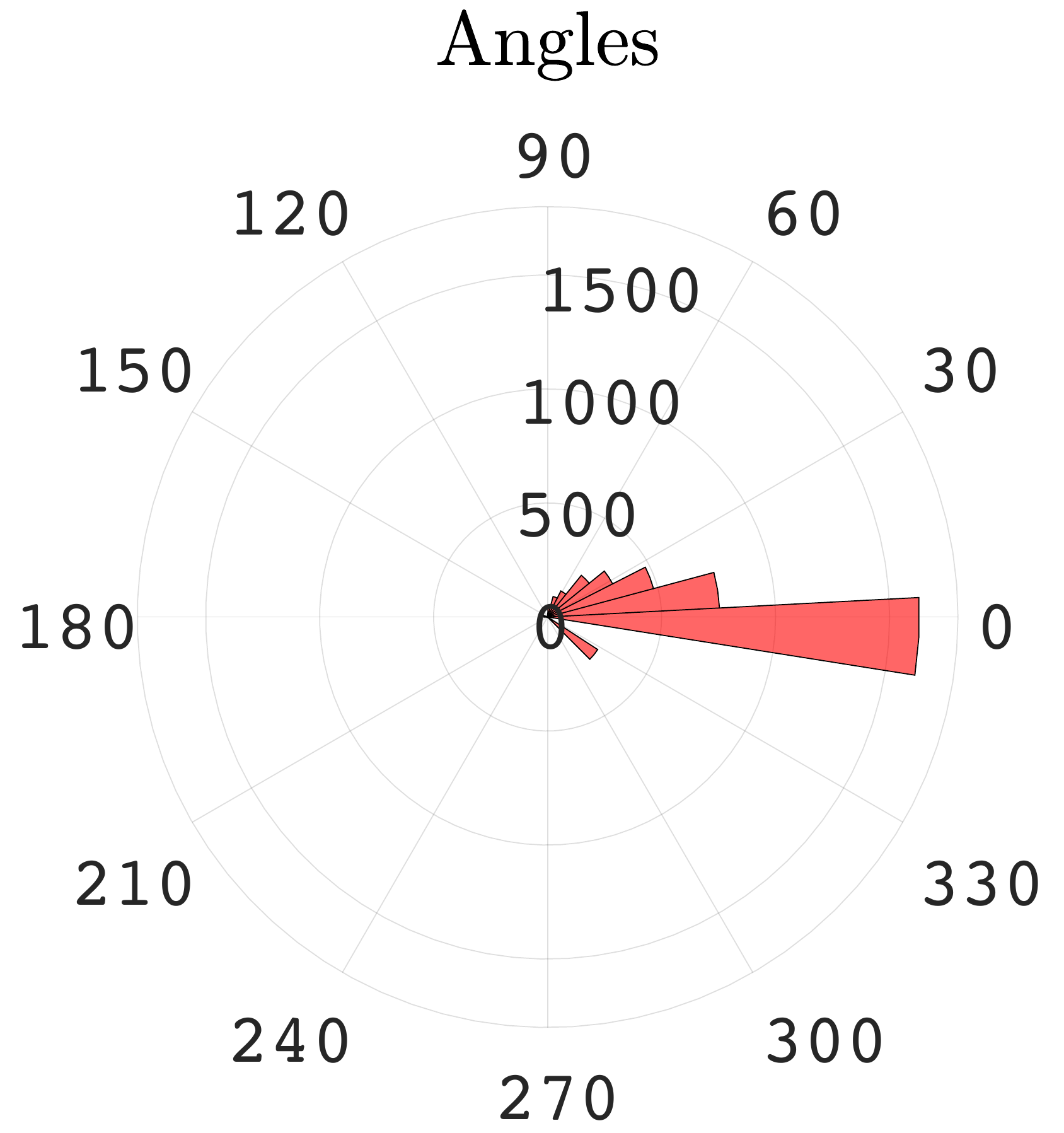}
    \caption{\gls{fsc}}
    \label{AOAMaxPowerFS}
    \end{subfigure}   
    \caption{Distribution of the \gls{aoa} associated with the \gls{mpc} with highest received power, \gls{nlos}}
        \label{fig:aoa-maxp}
\end{figure}

\subsection{Results}
\label{sec:results}

\begin{figure}[t]
    \centering
    \setlength\fwidth{.8\columnwidth}
    \setlength\fheight{.3\columnwidth}
\begin{tikzpicture}
\pgfplotsset{every tick label/.append style={font=\footnotesize}}

\definecolor{color0}{rgb}{0.75,0.75,0}

\begin{axis}[
width=\fwidth,
height=\fheight,
at={(0\fwidth,0\fheight)},
bar shift auto,
scale only axis,
legend cell align={left},
legend style={fill opacity=0.8, draw opacity=1, text opacity=1, draw=white!80!black, font=\footnotesize, at={(0.99, 0.99)}, anchor=north east},
legend columns=2,
tick align=inside,
tick pos=left,
x grid style={white!69.0196078431373!black},
xlabel={Channel Type},
xmin=-.5, xmax=2.5,
xtick style={color=black},
xtick=data,
xticklabels={TeraSim Channel,FS Channel,HB Channel},
y grid style={white!69.0196078431373!black},
ylabel={Latency [$\mu$s]},
ymin=0, ymax=105.000000,
ytick style={color=black},
ymajorgrids, xmajorgrids, yminorgrids,
xlabel style={font=\footnotesize\color{white!15!black}},
ylabel style={font=\footnotesize\color{white!15!black}}
]

\addplot [ybar, bar width=0.17, fill=red, draw=black, area legend]
  table[row sep=crcr]{%
0	3.031000\\
1	1.678000\\
2	1.678000\\
};
\addlegendentry{R = 4 Gbit/s}

\addplot [ybar, bar width=0.17, fill=blue, draw=black, area legend]
  table[row sep=crcr]{%
0	10.000000\\
1	1.678000\\
2	1.678000\\
};
\addlegendentry{R = 12 Gbit/s}

\addplot [ybar, bar width=0.17, fill=green, draw=black, area legend]
  table[row sep=crcr]{%
0	10.000000\\
1	2.278000\\
2	1.978000\\
};
\addlegendentry{R = 24 Gbit/s}

\addplot [ybar, bar width=0.17, fill=color0, draw=black, area legend]
  table[row sep=crcr]{%
0	100.000000\\
1	10.000000\\
2	2.267000\\
};
\addlegendentry{R = 60 Gbit/s}






\end{axis}

\end{tikzpicture}
    \caption{UDP latency with different channel models.}
    \label{channellatency}
\end{figure}
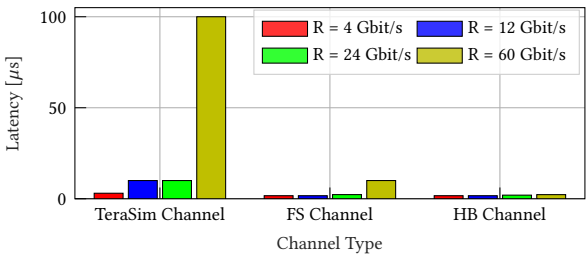

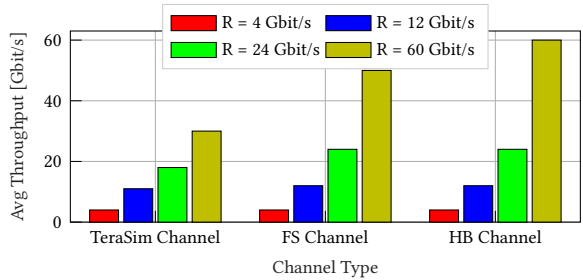
\begin{figure}[t]
    \centering
    \setlength\fwidth{.8\columnwidth}
    \setlength\fheight{.3\columnwidth}
\begin{tikzpicture}
\pgfplotsset{every tick label/.append style={font=\footnotesize}}

\definecolor{color0}{rgb}{0.75,0.75,0}

\begin{axis}[
width=\fwidth,
height=\fheight,
at={(0\fwidth,0\fheight)},
legend cell align={left},
legend style={
  fill opacity=1,
  draw opacity=1,
  text opacity=1,
  at={(0.5,0.8)},
  anchor=south,
  draw=white!80!black,
  font=\footnotesize
},
legend columns = 2,
tick align=inside,
tick pos=left,
scale only axis,
bar shift auto,
x grid style={white!69.0196078431373!black},
xlabel={Channel Type},
xmin=-0.5, xmax=2.5,
xtick style={color=black},
xtick=data,
xticklabels={TeraSim Channel,FS Channel,HB Channel},
y grid style={white!69.0196078431373!black},
ylabel={Avg Throughput [Gbit/s]},
ymin=0, ymax=63.000,
ytick style={color=black},
xlabel style={font=\footnotesize\color{white!15!black}},
ylabel style={font=\footnotesize\color{white!15!black}},
xmajorgrids, ymajorgrids, yminorgrids
]

\addplot [ybar, bar width=0.17, fill=red, draw=black, area legend]
  table[row sep=crcr]{%
0	4.000\\
1	4.000\\
2	4.000\\
};
\addlegendentry{R = 4 Gbit/s}

\addplot [ybar, bar width=0.17, fill=blue, draw=black, area legend]
  table[row sep=crcr]{%
0	11.000\\
1	12.000\\
2	12.000\\
};
\addlegendentry{R = 12 Gbit/s}

\addplot [ybar, bar width=0.17, fill=green, draw=black, area legend]
  table[row sep=crcr]{%
0	18.000\\
1	24.000\\
2	24.000\\
};
\addlegendentry{R = 24 Gbit/s}

\addplot [ybar, bar width=0.17, fill=color0, draw=black, area legend]
  table[row sep=crcr]{%
0	30.000\\
1	50.000\\
2	60.000\\
};
\addlegendentry{R = 60 Gbit/s}






\end{axis}

\end{tikzpicture}
    \caption{UDP throughput with different channel models.}
    \label{channelthroughput}
\end{figure}

Two sets of simulations are run to evaluate the performance of \gls{hbc} and \gls{fsc} channels based on the strongest received path. Figure~\ref{PowerCDF} shows the \gls{cdf} of the received power (including the antenna gain) for the strongest path of \gls{hbc} and \gls{fsc} channels for \gls{los} and \gls{nlos} scenarios. As can be seen, the \gls{hbc} model accounts for \glspl{mpc} with a higher received power in \gls{los}, with an offest smaller than 5 dB. The two channels are similar in \gls{nlos}, even though \gls{hbc} slightly outperfoms the \gls{fsc} in the third quartile. The spatial profile of the two channels is compared in Figure~\ref{fig:aoa}, which shows the distribution of the \gls{aoa} for the \gls{mpc} for both \gls{los} and \gls{nlos}. As expected, the \gls{los} channels are associated to fewer reflected \glspl{mpc}. Reflections play a minor (but more prominent) role also in the \gls{nlos} channels, where the direct path (even though attenuated by the presence of obstacles) is still the most common. This is shown also in Figure~\ref{fig:aoa-maxp}, which reports the distribution of the \gls{mpc} corresponding to the strongest path. In general, the \gls{hbc} channel models more \glspl{mpc} in the direction corresponding to \gls{los} than \gls{fsc}. This, as will be discussed later, has an impact when directional communications with extremely narrow beams are considered.


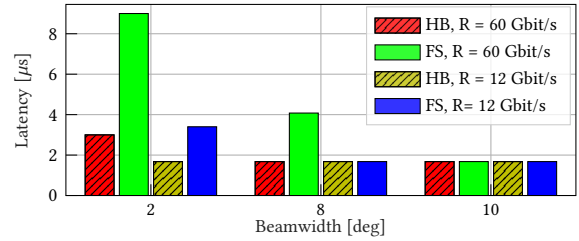
\begin{figure}[t]
    \centering
    \setlength\fwidth{.8\columnwidth}
    \setlength\fheight{.3\columnwidth}
\begin{tikzpicture}
\pgfplotsset{every tick label/.append style={font=\footnotesize}}

\definecolor{color0}{rgb}{0.75,0.75,0}

\begin{axis}[
width=\fwidth,
height=\fheight,
at={(0\fwidth,0\fheight)},
legend cell align={left},
legend style={
  fill opacity=0.8,
  draw opacity=1,
  text opacity=1,
  at={(0.99,0.99)},
  anchor=north east,
  draw=white!80!black,
  font=\footnotesize
},
legend columns = 1,
tick align=inside,
tick pos=left,
x grid style={white!69.0196078431373!black},
xlabel={Beamwidth [deg]},
xmin=-0.5, xmax=2.5,
scale only axis,
bar shift auto,
xtick style={color=black},
xtick=data,
xticklabels={2,8,10},
y grid style={white!69.0196078431373!black},
ylabel={Latency [$\mu$s]},
ymin=0, ymax=9.450000,
ytick style={color=black},
xlabel style={font=\footnotesize\color{white!15!black}},
ylabel style={font=\footnotesize\color{white!15!black}},
xmajorgrids, ymajorgrids, yminorgrids,
xlabel shift=-5pt
]

\addplot [ybar, bar width=0.17, fill=red, draw=black, area legend, postaction={pattern=north east lines}]
  table[row sep=crcr]{%
0	3.000000\\
1	1.678000\\
2	1.678000\\
};
\addlegendentry{HB, R = 60 Gbit/s}

\addplot [ybar, bar width=0.17, fill=green, draw=black, area legend]
  table[row sep=crcr]{%
0	9.000000\\
1	4.078000\\
2	1.678000\\
};
\addlegendentry{FS, R = 60 Gbit/s}

\addplot [ybar, bar width=0.17, fill=color0, draw=black, area legend, postaction={pattern=north east lines}]
  table[row sep=crcr]{%
0 1.678000\\
1 1.678000\\
2 1.678000\\
};
\addlegendentry{HB, R = 12 Gbit/s}

\addplot [ybar, bar width=0.17, fill=blue, draw=black, area legend]
  table[row sep=crcr]{%
0	3.400000\\
1	1.678000\\
2	1.678000\\
};
\addlegendentry{FS, R= 12 Gbit/s}





\end{axis}

\end{tikzpicture}
    \caption{UDP latency for different RX beamwidth.}
    \label{channellatencybeam}
\end{figure}

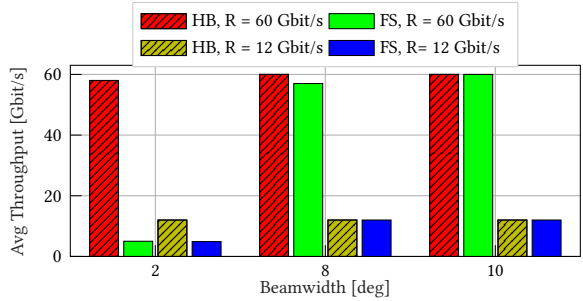
\begin{figure}[t]
    \centering
    \setlength\fwidth{.8\columnwidth}
    \setlength\fheight{.3\columnwidth}
\begin{tikzpicture}
\pgfplotsset{every tick label/.append style={font=\footnotesize}}

\definecolor{color0}{rgb}{0.75,0.75,0}

\begin{axis}[
width=\fwidth,
height=\fheight,
at={(0\fwidth,0\fheight)},
legend cell align={left},
legend style={
  fill opacity=1,
  draw opacity=1,
  text opacity=1,
  at={(0.5,0.99)},
  anchor=south,
  draw=white!80!black,
  font=\footnotesize
},
legend columns = 2,
tick align=inside,
tick pos=left,
x grid style={white!69.0196078431373!black},
xlabel={Beamwidth [deg]},
xmin=-0.5, xmax=2.5,
scale only axis,
bar shift auto,
xtick style={color=black},
xtick=data,
xticklabels={2,8,10},
y grid style={white!69.0196078431373!black},
ylabel={Avg Throughput [Gbit/s]},
ymin=0, ymax=63.000,
ytick style={color=black},
xlabel style={font=\footnotesize\color{white!15!black}},
ylabel style={font=\footnotesize\color{white!15!black}},
xmajorgrids, ymajorgrids, yminorgrids,
xlabel shift=-5pt
]

\addplot [ybar, bar width=0.17, fill=red, draw=black, area legend, postaction={pattern=north east lines}]
  table[row sep=crcr]{%
0	58.000\\
1	60.000\\
2	60.000\\
};
\addlegendentry{HB, R = 60 Gbit/s}

\addplot [ybar, bar width=0.17, fill=green, draw=black, area legend]
  table[row sep=crcr]{%
0	5.000\\
1	57.000\\
2	60.000\\
};
\addlegendentry{FS, R = 60 Gbit/s}

\addplot [ybar, bar width=0.17, fill=color0, draw=black, area legend, postaction={pattern=north east lines}]
  table[row sep=crcr]{%
0 12.000\\
1 12.000\\
2 12.000\\
};
\addlegendentry{HB, R = 12 Gbit/s}

\addplot [ybar, bar width=0.17, fill=blue, draw=black, area legend]
  table[row sep=crcr]{%
0	4.900\\
1	12.000\\
2	12.000\\
};
\addlegendentry{FS, R= 12 Gbit/s}





\end{axis}

\end{tikzpicture}
    \caption{UDP throughput for different RX beamwidth.}
    \label{channelthroughputbeam}
\end{figure}

Figures~\ref{channellatency} and~\ref{channelthroughput} compare the latency and throughput, respectively, with the default TeraSim channel model (\gls{los} only), and the \gls{hbc} and \gls{fsc} channels with different source rate values. For the smaller source rate values, the performance with the three channel models is similar, with a slightly higher latency with the TeraSim channel. Indeed, as the bandwidth is underutilized, more robust \glspl{mcs} and buffering can compensate for a weaker channel or a suboptimal coordination among the nodes for directional communications. When the source rate increases to 60 Gbit/s, the difference in the performance measured with the three models increases, with the \gls{hbc} model associated to a higher throughput, and, consequently, lower latency. Notably, a throughput comparable with the full source rate can be achieved only in simulations based on the \gls{hbc} model. This can be explained by analyzing the interaction between the rotating directional antenna model (described in Sec.~\ref{sec:sim}) and the \gls{mpc} \gls{aoa} distribution. The \gls{hbc} channel has a higher probability of having the strongest \gls{nlos} \gls{mpc} given by the attenuated component in the \gls{los} direction, thus it is less likely to require adaptation of the antenna direction at the \gls{tx} and \gls{rx}. This results into a higher received power than with \gls{fsc}, and thus a higher throughput.



This consideration extends to the results shown in Figures~\ref{channellatencybeam} and~\ref{channelthroughputbeam}, which report the latency and throughput for different values of the \gls{rx} antenna beamwidth and the \gls{hbc} and \gls{fsc} channels. Even for the highest source rate, the interaction between the directional antenna model with larger beamwidths (e.g., 10 degrees) and the spatial distribution of the \glspl{mpc} does not result into different throughput or latency for the \gls{hbc} and \gls{fsc} channel models. When smaller beamwidths are used (i.e., 2 degrees), there is instead a difference in performance between the \gls{hbc} and \gls{fsc} models for both source rate values, showing that the modeling of the interaction between the antenna model and the channel model has an impact on the full stack performance.

\begin{figure}[t]
    \centering
    \setlength\fwidth{\columnwidth}
    \setlength\fheight{.6\columnwidth}
    \input{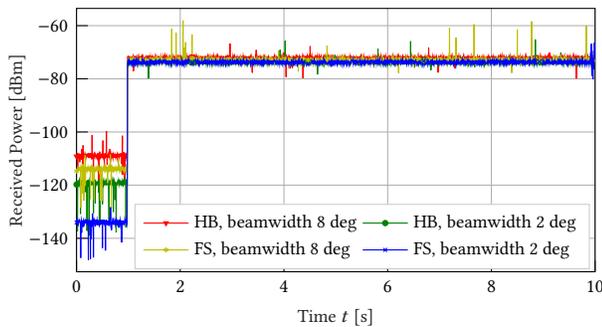}
    \setlength\abovecaptionskip{.1cm}
    \caption{Received power with different channel models and beams.}
    \label{fig:receivedpowerbeam}
\end{figure}

Finally, we analyze the effect of beamwidth on the received power. In particular, Figure~\ref{fig:receivedpowerbeam} reports the evolution of the received power vs. time, with users following the mobility pattern of Figure~\ref{Scenario}, for a single simulation run. For this figure, the \gls{nlos}/\gls{los} transition of the \gls{fsc} channel is not random, but it follows that modeled through the \gls{rt} for the specific simulation scenario. In \gls{nlos} (i.e., before $t=1.5$ s), there is a significant difference between the different channel models, with an average gap of 10 dB between the two beamwidth values for \gls{hbc}, and of 25 dB for \gls{fsc}. 
The different beamwidths and channel models have a more limited impact in the \gls{los} scenario, i.e., after $t=1.5$ s, due to the direct link between \gls{rx} and \gls{tx}. The results of Figure~\ref{fig:receivedpowerbeam} also highlight one shortcoming of currently available models for above 100 GHz, i.e., the lack of a temporal characterization of the \gls{los}/\gls{nlos} transitions. 


\section{Conclusions}
\label{sec:conclusions}

This paper introduced the implementation of two \glspl{scm} for the ns-3 module TeraSim, based on the 140 GHz models from~\cite{Statchannel,chen2021channel}. This makes it possible to simulate \gls{los} and \gls{nlos} scenarios for future \gls{6g} networks, including fading and the possibility to interact with antenna array models. We also compared the two channel models (based on a fully stochastic and an \gls{rt}-based modeling) with simulations in an indoor scenario, which have highlighted how the two modeling strategies differ in their interaction with the directional antenna model of TeraSim.

As future work, we will further expand the analysis of the full-stack performance, including different source traffic patterns, scenarios, and different protocol stack implementations. 

\begin{acks}
This work is supported in part by the EU MSCA ITN project MINTS
“MIllimeter-wave NeTworking and Sensing for Beyond 5G” (grant no. 861222).
\end{acks}

\bibliographystyle{ACM-Reference-Format.bst}
\bibliography{bibl.bib}

\end{document}